\documentclass[sigconf]{acmart}

\AtBeginDocument{%
  }

\setcopyright{acmlicensed}
\copyrightyear{2026}
\acmYear{2026}
\acmDOI{XXXXXXX.XXXXXXX}
\acmConference[FORGE '26]{The 3rd ACM International Conference on AI Foundation Models and Software Engineering}{April 12--13, 2026}{Rio de Janeiro, Brazil}

\usepackage{tabularx}
\usepackage{paralist}
\usepackage{booktabs}
\usepackage{multirow}
\usepackage{tcolorbox}
\tcbuselibrary{theorems}

\newcommand{\sectopic}[1]{\vspace{0.2em}\par\noindent{\textit{\bfseries #1}}}

\newtcbtheorem{system}{System Prompt}
{sharp corners=all,colback=gray!5,colframe=blue!35!black,fonttitle=\bfseries,left=0.1cm,right=0.1cm,top=0.1cm,bottom=0.1cm}{th}

\begin{document}

\title[Reporting LLM Prompting in Automated SE: A Guideline Based on Current Practices and Expectations]{Reporting LLM Prompting in Automated Software Engineering: \\ A Guideline Based on Current Practices and Expectations}


\author{Alexander Korn}
\email{alexander.korn@uni-due.de}
\affiliation{%
  \institution{University of Duisburg-Essen}
  \city{Essen}
  \country{Germany}
}

\author{Lea Zaruchas}
\affiliation{%
  \institution{University of Cologne}
  \city{Cologne}
  \country{Germany}
}

\author{Chetan Arora}
\email{chetan.arora@monash.edu}
\affiliation{%
  \institution{Monash University}
  \city{Melbourne}
  \country{Australia}
}

\author{Andreas Metzger}
\email{andreas.metzger@uni-due.de}
\affiliation{%
  \institution{University of Duisburg-Essen}
  \city{Essen}
  \country{Germany}
}

\author{Sven Smolka}
\email{sven.smolka@uni-due.de}
\affiliation{%
  \institution{University of Duisburg-Essen}
  \city{Essen}
  \country{Germany}
}

\author{Fanyu Wang}
\email{fanyu.wang@monash.edu}
\affiliation{%
  \institution{Monash University}
  \city{Melbourne}
  \country{Australia}
}

\author{Andreas Vogelsang}
\email{andreas.vogelsang@uni-due.de}
\affiliation{%
  \institution{University of Duisburg-Essen}
  \city{Essen}
  \country{Germany}
}

\renewcommand{\shortauthors}{Korn et al.}

\begin{abstract}
Large Language Models, particularly decoder-only generative models such as GPT, are increasingly used to automate Software Engineering tasks. These models are primarily guided through natural language prompts, making prompt engineering a critical factor in system performance and behavior. Despite their growing role in SE research, prompt-related decisions are rarely documented in a systematic or transparent manner, hindering reproducibility and comparability across studies.
To address this gap, we conducted a two-phase empirical study. First, we analyzed nearly 300 papers published at the top-3 SE conferences since 2022 to assess how prompt design, testing, and optimization are currently reported. Second, we surveyed 105 program committee members from these conferences to capture their expectations for prompt reporting in LLM-driven research. Based on the findings, we derived a structured guideline that distinguishes essential, desirable, and exceptional reporting elements.
Our results reveal significant misalignment between current practices and reviewer expectations, particularly regarding version disclosure, prompt justification, and threats to validity. We present our guideline as a step toward improving transparency, reproducibility, and methodological rigor in LLM-based SE research.
\end{abstract}




\maketitle

\section{Introduction}

Automated software engineering is about applying computational methods and tools to automate various activities within the software engineering life cycle. 
Large Language Models (LLMs), particularly decoder-only generative models such as GPT-3~\cite{brown2020language}, have rapidly transformed how Software Engineering (SE) tasks can be automated~\cite{hou2024large}. 
These models simultaneously ``speak'' fluent natural language and multiple programming languages, making them attractive for several SE tasks, such as code synthesis, test generation, defect repair, requirements analysis, and beyond~\cite{nguyen2023generative}. LLMs are primarily guided by textual prompts. Prompting, i.e., giving carefully designed textual instructions, has become the primary interface to guide model behavior, making \textit{Prompt Engineering (PE)} a crucial factor in LLM performance~\cite{liu2023pre-train,Shin2025,Sahoo24}.

The use of decoder-only LLMs in SE research has steadily increased recently. As we will show later in this paper, we identified almost 300 papers published in the top three SE conferences since 2022 that have leveraged such models to automate a wide range of SE tasks. Yet, despite their widespread use, how prompts are constructed, refined, and evaluated is rarely reported in a systematic or transparent manner.

Despite the central role of prompting in determining model behavior, current research practices lack consistency in how prompts are documented and justified, as expected of other SE experimental artifacts. There is no established standard for reporting prompt design, testing, or optimization. As a result, prompt-related decisions are often underreported or omitted entirely, limiting reproducibility, reducing comparability between studies, and ultimately hindering progress in this fast-moving domain.

Given the novelty and evolving nature of LLM-based SE research, it is premature to impose top-down standards based solely on expert opinion. Instead, we argue that reporting guidelines should emerge from observed practices and the expectations of the research community itself. Understanding how prompts are currently reported and how they should be requires empirical investigation grounded in both literature analysis and researcher insight.

To address this need, we conducted a multi-phase empirical study structured around the following research questions:

\textbf{RQ1: How do researchers currently report prompts in SE research papers?}
     We analyzed nearly 300 papers published in ICSE, FSE, and ASE since 2022 to assess how authors report on prompt design, validation, and optimization.
     
\textbf{RQ2: What are the expectations of SE researchers regarding prompt creation, evaluation, and reporting in SE research papers?}
     We surveyed 105 Program Committee (PC) members of the aforementioned conferences to capture their expectations for prompt reporting practices.
     
\textbf{RQ3: How consistent is the current state with the expectations?}
     We distilled the expressed expectations into a guideline and compared them against current reporting practices.

This work makes the following contributions:

\begin{compactenum}
    \item A taxonomy and frequency analysis of how prompt design, testing, and optimization are currently reported in nearly 300 SE research papers (RQ1).
    \item An empirically derived set of reporting expectations from SE reviewers (RQ2).
    \item A comparative analysis revealing gaps and alignments between current practices and community expectations (RQ3).
    \item A guideline grounded in empirical evidence to enhance transparency, reproducibility, and comparability in LLM-based SE research.
\end{compactenum}

By synthesizing current practices and reviewer expectations, this study aims to raise the methodological standard for LLM-based research in software engineering and to support future work with more transparent and reproducible foundations.

\section*{Data Availability}
All data we used in our study and the code used to analyze the data are available in our replication package\footnote{\url{https://doi.org/10.5281/zenodo.16101751}}.

\section{Related Work}
Recent Systematic Literature Reviews (SLRs) provide an analysis of the use of LLMs for automating SE tasks, aiming to determine which LLMs are used, the methods for data collection and preparation, the strategies for prompt engineering, and the techniques for optimizing and evaluating the performance of LLMs. Several such SLRs focus on specific SE tasks, such as requirements engineering~\cite{huang2025}, code generation~\cite{ZanCZLWGWL23}, program repair~\cite{ZhangFMSC24}, and software testing~\cite{WangHCLWW24}. In contrast, Hou~et~al.~\cite{hou2024large} performed an SLR covering the entire software development life cycle.

While the aforementioned publications provide important insights into how LLMs are used in automated software engineering, they do not offer explicit guidelines on how prompting should be reported in SE research.

Trinkenreich et al.~\cite{TrinkenreichEtAl2025} issue a call to action for the SE research community to develop reporting guidelines to ensure continued rigor and impact. In particular, they advocate for transparent reporting when using LLMs, including specification of the model and version, prompting strategies, and mechanisms for human oversight. However, their work stops short of proposing concrete, operationalized guidelines.

Baltes et al.~\cite{baltes2025guidelines} propose a set of guidelines for the use and evaluation of LLMs in SE. Developed through expert discussions at the 2024 International Software Engineering Research Network (ISERN) meeting, their guidelines follow a top-down consensus-driven process and cover a wide range of topics, including tool architecture, evaluation metrics, baselines, and benchmarks. For prompt reporting specifically, they define eight recommendations, five marked as \textit{MUST} and three as \textit{SHOULD}, which call for full prompt disclosure (including structure and formatting), rationale for prompt design, reuse documentation, input handling, and interaction log sharing. The guidelines are available in an open source repository\footnote{\url{https://llm-guidelines.org/}}.

Our work complements this effort by empirically assessing the current state of prompt reporting in SE research and capturing the expectations of PC members. While Baltes~et~al.'s guidelines reflect expert consensus, our guidelines are derived bottom-up from observed reporting practices in nearly 300 SE research papers and validated through a survey of PC members from top-tier SE conferences. 

In addition, we see our findings as a valuable contribution to ongoing community efforts aimed at standardizing empirical research practices. In particular, the ACM SIGSOFT Empirical Standards\footnote{\url{https://www2.sigsoft.org/EmpiricalStandards/}} initiative currently provides structured guidance on study design and reporting across various empirical methods but does not yet include standards tailored to LLM-based research. By contributing empirically grounded, task-specific insights into prompt reporting, we aim to help fill this gap and support the development of future standards for the transparent and reproducible use of LLMs in SE.



\section{Current State of Prompt Reporting (RQ1)}
\label{sec:rq1}

The goal of this RQ is to analyze how prompting is described and evaluated in recent research papers proposing LLM-driven SE approaches.

To investigate how prompts are currently reported in software engineering research, we conducted a systematic review of recent publications from leading conferences. The goal of this review was to assess the extent, consistency, and depth of prompt reporting practices across empirical studies involving large language models.

We selected a literature review approach to obtain an objective and comprehensive understanding of the current state of practice. Systematic reviews are a well-established method for synthesizing research evidence and are particularly effective for identifying trends, gaps, and variations in how specific techniques or artifacts, such as prompts, are used and reported in published work~\cite{kitchenham2007guidelines}.

\subsection{Study Design}
We answer RQ1 through a SLR~\cite{kitchenham2002principles}. By examining publications from top-tier SE conferences, we aim to identify common practices in prompt documentation, the level of detail provided, and the reported techniques used to create and evaluate prompts. The findings will highlight potential reporting gaps, assess inconsistencies between studies, and provide information on how PE is currently approached in SE research. This review will serve as a foundation for understanding the state of prompt reporting and will contribute to establishing best practices for future studies.

\sectopic{Paper selection:} We began our paper selection process by collecting all papers published in 2022 or later from the three top SE conferences according to the CORE ranking\footnote{\url{https://portal.core.edu.au/conf-ranks/}}: the IEEE/ACM International Conference on Software Engineering (ICSE'22--ICSE'25), the ACM International Conference on the Foundations of Software Engineering (FSE'22--FSE'24), and the IEEE/ACM International Conference on Automated Software Engineering (ASE'22--ASE'24).\footnote{The proceedings of FSE'25 appeared too shortly before the submission deadline to be included.} We chose 2022 as the starting year based on the assumption that the use of decoder-only LLMs was rare before the release of ChatGPT in December 2021.

We limited our scope to these conferences to focus on venues that typically reflect the highest methodological standards and most up-to-date research practices. Journal papers were excluded to maintain a consistent corpus, as their extended length and format may lead to substantially different reporting behaviors compared to page-restricted conference papers. Moreover, given their longer review cycles, many journal articles may not yet have been relevant at the time of our analysis.

We filtered these papers in three steps to ensure that only those relevant to our study were retained in the dataset.

\begin{compactenum}
    \item \textit{Filtering by document length:} We considered only full papers.
    If a paper has fewer than 7 pages, we exclude it because short papers may not present fully developed approaches, preliminary evaluations, or make compromises due to page limitations.
    \item \textit{Filtering by keywords:} To filter irrelevant papers, we defined a set of keywords that we expected to be present in any relevant paper. Papers not including any of the following keywords were excluded: \textit{LLM, LLMs, Large Language Model, GenAI, Generative AI, OpenAI, GPT, ChatGPT, Llama, Claude, Prompt, Prompting, Prompted}.
    \item \textit{LLM-based filtering:} As the final step, we used different LLMs to further filter the papers. 
    We prompted the LLMs to include only papers that a) focus primarily on automating SE tasks, b) use generative LLMs (i.e., decoder-only models), and c) conduct primary studies (i.e., no meta-analyses, literature reviews, etc.). 
    The full prompt is given below.
\end{compactenum}

\begin{system*}{LLM-Based Filtering}
You are a researcher conducting a literature review. You will be given the full text of various academic papers.
Your task is to decide whether each paper should be included based on the following strict criteria: \\
Inclusion Criteria:
\begin{itemize}
    \item The primary focus of the study must be on automating software engineering (SE) tasks.
    \item The study must utilize generative LLMs (decoder-only architectures, e.g., GPT models).
    \item The study must be a primary study (i.e., proposing, evaluating, or implementing a method). Meta-analyses, literature reviews, and systematic reviews must be excluded.
\end{itemize}

Instructions:
\begin{itemize}
    \item Base your decision solely on the content of the paper.
    \item If the paper does not clearly meet all criteria, exclude it.
    \item Respond with exactly one word: "include" or "exclude".
    \item Do not provide explanations, justifications, or additional text.
\end{itemize}

Examples: \\
\{four examples including a paper's title, a one-sentence summary, and whether to include or exclude that paper\}
\end{system*}

We tested different prompts for the LLM-based filtering, employing PE techniques such as \textit{role prompting}, \textit{zero-shot prompting}, and \textit{few-shot prompting}. After testing various prompts across multiple models, we selected the final prompt based on its ability to achieve the highest recall by correctly including papers that met the inclusion criteria. We evaluated this approach by manually screening all papers from ICSE'22 to ICSE'24 using the same inclusion criteria provided to the LLM in the prompt. We then compared the LLM-based filtering results to the outcomes of the manual screening.

For the final filtering process, we used three different LLMs (\textit{gpt-4.1-mini-2025-04-14}, \textit{deepseek-v3-0324}, and \textit{gemini-2.5-flash-preview-05-20}), accessing them via their respective APIs. For all models, we maintained a temperature of $1.0$. We also tested lower temperature settings, which did not lead to improved results. We did not configure any other settings.
We selected the three models because (a) they achieved the best performance in our comparative tests with other models, (b) they were the most cost-effective out of the tested models, and (c) they are offered by different providers, which we considered beneficial for enhancing the trustworthiness of the approach by mitigating potential provider-specific biases. The precision and recall metrics for these models are reported in~\autoref{tab:llm-based-filtering-results}.

The best-performing model, \textit{gpt-4.1-mini-2025-04-14}, achieved a recall of 84.83~\%. For the final filtering step, we included a paper if any of the three models recommended its inclusion. This strategy allowed us to reach a combined recall of 98.28~\%, which we deemed sufficient given the substantial reduction in manual workload enabled by pre-filtering. The approach prioritized maintaining high recall despite the risk of decreasing precision, since manual data extraction was still conducted afterward, allowing us to identify and exclude any false positives at a later stage.

%
Table~\ref{tbl:paper-selection} provides an overview of the number of papers before and after each filtering step.


\begin{table}
    \centering
    \caption{LLM-Based Filtering Performance}
    \newcolumntype{R}{>{\raggedleft\arraybackslash}X}
    \begin{tabularx}{\linewidth}{@{}p{4.2cm} RR@{}}
        \toprule
         \textbf{Model} & \textbf{Precision} & \textbf{Recall} \\ \midrule
         gpt-4.1-mini-2025-04-14 & 68.75~\% & 94.83~\% \\
         deepseek-v3-0324 & 88.46~\% & 79.31~\% \\
         gemini-2.5-flash-preview-05-20 & 86.54~\% & 77.59~\% \\
         \textbf{Combined result ($\geq$ 1 of 3)} & \textbf{67.86~\%} & \textbf{98.28~\%} \\
         \bottomrule
    \end{tabularx}
    \label{tab:llm-based-filtering-results}
\end{table}

\begin{table*}
    \centering
    \caption{Paper Selection Process}
    \label{tbl:paper-selection}
    \newcolumntype{R}{>{\raggedleft\arraybackslash}X}
    \begin{tabularx}{\textwidth}{p{6cm} RRRR RRR RRR R}
        \toprule
        & \multicolumn{4}{c}{\textbf{ICSE}} & \multicolumn{3}{c}{\textbf{FSE}} & \multicolumn{3}{c}{\textbf{ASE}} & \multirow{2}{*}{\textbf{Sum}} \\
        \cmidrule(lr){2-5}\cmidrule(lr){6-8}\cmidrule(lr){9-11}
        & \textbf{2022} & \textbf{2023} & \textbf{2024} & \textbf{2025} & \textbf{2022} & \textbf{2023} & \textbf{2024} & \textbf{2022} & \textbf{2023} & \textbf{2024} & \\
        \midrule
        \# of published papers & 197 & 211 & 237 & 246 & 186 & 205 & 121 & 213 & 209 & 265 & \textbf{2,090} \\
        \hspace{0.55em}\# of full papers ($\geq$ 7 pages) & 197 & 207 & 234 & 246 & 130 & 161 & 121 & 116 & 145 & 174 & \textbf{1,731} \\
        \hspace{1.1em}\# of full papers with keywords & 24 & 44 & 94 & 152 & 19 & 55 & 61 & 23 & 74 & 138 & \textbf{684} \\
        \hspace{1.65em}\# of full papers after LLM-based filtering & 5 & 13 & 40 & 97 & 7 & 20 & 42 & 7 & 34 & 67 & \textbf{332} \\
        \textbf{final \# of papers after manual analysis} & \textbf{4} & \textbf{11} & \textbf{38} & \textbf{91} & \textbf{3} & \textbf{10} & \textbf{33} & \textbf{3} & \textbf{22} & 71 & \textbf{286} \\
        \bottomrule
    \end{tabularx}
\end{table*}

\sectopic{Data extraction:} After selecting the relevant papers, we proceeded with the data extraction. We created an extraction sheet containing 11 questions, which are listed in \autoref{tbl:data-extraction}. A more detailed description of the questions, including explicit instructions on how to answer them, is provided in the extraction sheet, which is part of our replication package.  

\begin{table}
    \centering
    \caption{Data Extraction Sheet. 
    C = Closed-ended question (yes/no/partially), O = Open-ended question}
    \label{tbl:data-extraction}
    \begin{tabularx}{\columnwidth}{@{}lXc@{}}
        \toprule
        \textbf{ID} & \textbf{Question} & \textbf{C/O} \\
        \midrule
        \multicolumn{3}{@{}l@{}}{\textit{1. LLM Usage}} \\ \midrule
        E1 & Which LLM(s) was/were used? & O \\
        E2 & Which configuration parameters of the LLM(s) are reported? & O \\ \midrule
        \multicolumn{2}{@{}l@{}}{\textit{2. Prompt Description and Design}} \\ \midrule
        E3 & Is the full prompt provided word-by-word? & C \\
        & \textit{Does the paper provide the full prompt(s) word-by-word, e.g. in a listing? Placeholders and templates are allowed.} \\
        E4 & Is the prompt and its structure explained? & C \\
        & \textit{Does the paper explain the prompt and its structure, e.g., by describing it in the text outside of the word-by-word representation?} \\
        E5 & Do the authors justify why they constructed the prompt the way they did? & C \\
        & \textit{What is the rationale for choosing a certain PE technique, structure, phrasing, context, etc.?} \\
        E6 & Which PE techniques are reported? & O \\ \midrule
        \multicolumn{2}{@{}l@{}}{\textit{3. Prompt Testing and Evaluation}} \\ \midrule
        E7 & Does the paper mention any form of prompt testing or tuning (e.g., prompt refinement) to improve LLM performance? & C \\
        & \textit{This question focuses on whether the paper mentions testing or tuning prompts without needing to provide specific details.} \\
        E8 & Does the paper report results of multiple prompt variations? & C\\
        & \textit{This question examines whether the paper explicitly describes variations in prompts, provides details on how they differ, and presents results for those variations.} \\ \midrule
        \multicolumn{2}{@{}l@{}}{\textit{4. Threats to Validity}} \\ \midrule
        E9 & Is prompting seen as a threat to validity? & C \\ \midrule
        \multicolumn{2}{@{}l@{}}{\textit{5. Software Engineering Tasks}} \\ \midrule
        E10 & For which task categories was/were the LLM(s) used? & O \\
        E11 & For which tasks was/were the LLM(s) used? & O \\
        \bottomrule
    \end{tabularx}
\end{table}

Questions E3--E5 and E7--E9 were closed-ended, allowing only \textit{yes}, \textit{no}, or \textit{partially} as possible answers. The option \textit{partially} was permitted only when multiple prompts or LLMs were used in the paper, but the question could not be answered consistently for all of them. While question E1 was a free-text question, questions E2, E6, E10, and E11 were open-text questions constrained by a predefined set of answers developed by the authors of this paper during extraction. Question E10 specifically consisted only of \textit{software development life cycle (SDLC)} phases, while question E11 initially used all tasks extracted by Hou et al.~\cite{hou2024large}, enabling comparability to their study.

To ensure consistency in data extraction among the six authors of this paper, we conducted three extraction rounds, with the first two serving as test rounds. In the first round, each author extracted data from 10 papers. The papers were assigned with an overlap such that each paper was reviewed by two authors, and each author's set overlapped with those of two other authors. After extraction, we manually examined the differences, discussed misunderstandings in the extraction sheet, and refined the questions to improve clarity.

Following these first improvements, we conducted a second round in which each author again received 10 papers with overlapping assignments similar to round 1. Again, data were extracted independently, and any remaining misunderstandings were discussed to finalize the extraction sheet and align everyone's understanding of the questions to ensure consistent data extraction.

For the final round of extraction, we divided all remaining papers from the filtering step (332; cf. \autoref{tbl:paper-selection}) equally among the authors. This time, there was no overlap, with each author reviewing a unique subset of the papers. While this approach was chosen for time efficiency, we were confident in its validity given the prior two rounds, which served to harmonize our understanding of the extraction process. Additionally, two authors cross-validated a random sample of 5 papers of the other raters to uncover any remaining systematic inconsistencies.

\subsection{Study Results}
\begin{figure}
    \centering
    \includegraphics[width=\columnwidth]{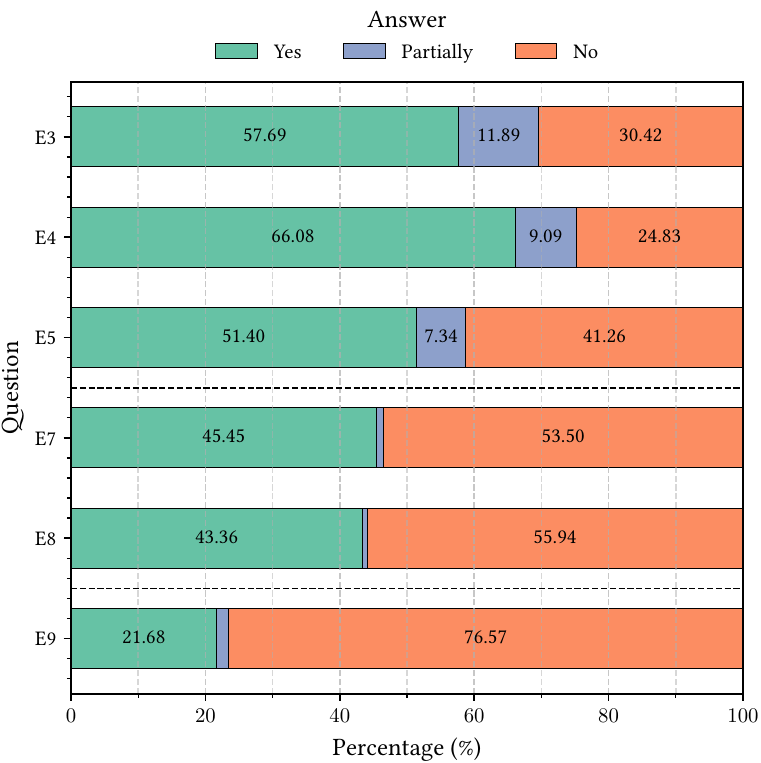}
    \caption{A bar chart showing the percentages of closed-ended extraction questions (cf. \autoref{tbl:data-extraction}) answered with either \textit{Yes}, \textit{No}, or \textit{Partially}.}
    \label{fig:extraction-closed-questions}
\end{figure}

The results represent extracted data from the final list of 286 papers (see \autoref{tbl:paper-selection}).
\autoref{fig:extraction-closed-questions} shows an overview of the results for the closed-ended questions. Together with these results, we report the results of the open-ended questions in the following. 

\sectopic{LLM usage:} \textit{E1 (used LLM):} Most papers (92.31~\%) mention the name of the LLM used in their study. In 39.16~\% of the papers, the number of parameters is included as part of the name (e.g., \textit{llama3 70b}). The exact version is specified in only 16.43~\% of the papers. We did not count labels such as \textit{gpt-3.5} as specifying a version, since GPT and other models can vary significantly for different major versions, effectively making them separate models. Instead, we treated specific snapshots or dates as indicating a version (e.g., \textit{0125} or \textit{2024-05-13}). The most used models were \textit{gpt-3.5-turbo} (63 instances), \textit{gpt-4} (61 instances), \textit{codellama} (36 instances), \textit{gpt-3.5} (27 instances), and \textit{text-davinci-003} (12 instances).

\textit{E2 (configuration parameters):} Of all papers, 69.93~\% reported at least one configuration parameter. The most commonly reported parameters were the temperature (131 instances), output token limit (33 instances), top-\textit{p} value (29 instances), number of prompt iterations (24 instances), and input token limit (23 instances).

\sectopic{Prompt description and design:} 
\textit{E3, E4, E5 (prompt documentation):} In a majority of papers, the authors either fully or partially describe the used prompt(s) and their structure (75.17~\%; yes + partially). In more than half of all papers, the authors even provide the used prompt(s) word-by-word (69.58~\%). Some authors provided detailed descriptions of the prompts without listing the exact wording, leading to higher positive responses for question E4. Around half of the papers (58.74~\%) specifically justified their prompt construction, i.e., they gave reasons for why they created the prompt in the specified way.

\textit{E6 (PE techniques):} In 62.24~\% of the papers, the authors report which PE techniques they use. The most frequently reported PE techniques were \textit{few-shot prompting} (62 instances), \textit{chain-of-thought prompting} (53 instances), \textit{zero-shot prompting} (49 instances), \textit{in-context learning} (27 instances), and \textit{retrieval-augmented generation} (19 instances). A total of 50 unique PE techniques were mentioned across all papers. The full list is included in our replication package. It is important to note that we extracted PE techniques only if they were explicitly mentioned as such by the authors, i.e., we did not identify PE techniques by ourselves (e.g., by analyzing the prompts).

\sectopic{Prompt testing and evaluation:} 
\textit{E7 (Prompt tuning):} In only 46.5~\% of papers, the authors fully or partially mention that they refined or tuned the used prompts as part of their research process. In the remaining 53.5~\% of papers, there is no indication that the authors have refined the prompts during their research process. 
Automated prompt-tuning techniques, such as \textit{self-refinement}, were used only rarely (in 4.9~\% of papers).

\textit{E8 (Prompt comparison):}
In 44.06~\% of the papers, the authors explicitly describe different prompt variations and also provide results of their performance.

\sectopic{Prompting as a threat to validity:} 
\textit{E9:}
Only 23.43~\% of the papers explicitly report prompting as part of their threats to validity. In these papers, the authors usually mention that the results may be influenced by the composition and phrasing of prompts. Rephrasing or optimizing the prompts may change the results.  

\subsection{Threats to Validity}

\sectopic{Construct Validity:}  
Our analysis focuses exclusively on papers from the top-3 SE conferences (ICSE, FSE, ASE), which may not fully represent the broader SE research landscape. While these venues are highly selective and influential, important LLM-based work may appear in other venues, such as specialized conferences or journals.  
Additionally, publication bias may affect our results: papers that successfully applied prompting may be more likely to be accepted and published, potentially skewing the picture of actual practices.  
Furthermore, our corpus includes papers published up to mid-2025. Due to conference submission timelines, many of these papers were likely written no later than mid-2024. As a result, our findings may lag behind current practices or recent trends in prompt reporting.

\sectopic{Internal Validity:}  
Despite systematic procedures, there are risks of bias in paper selection and data extraction. Although we used keyword-based and LLM-assisted filtering to identify relevant papers, it is possible that some relevant studies were unintentionally excluded. 
We tested several prompts used to filter papers automatically and finally ended up with a prompt that achieved a high recall. However, further prompt tuning may achieve even better recall. 
Moreover, although we employed a method for information extraction that allowed selection only from an extendable predefined list, the interpretation of reporting practices may still involve subjective judgment, particularly in borderline cases or when details were ambiguous. We mitigated this issue by cross-checking coding among researchers; however, subjectivity cannot be fully eliminated.

\sectopic{External Validity:}  
Our findings may not generalize to all SE research involving LLMs, especially in industry or non-academic settings, where PE practices and documentation norms may differ significantly.  
Additionally, practices in other research communities that use LLMs, such as NLP, HCI, or education, might follow different reporting standards. Thus, while our findings are grounded in the SE community, the proposed guidelines may not directly apply to other domains.  
Finally, as the LLM ecosystem evolves rapidly, our results may become outdated as tools, APIs, and community norms change, potentially limiting the long-term applicability of our analysis.

\section{Expectations of SE Researchers (RQ2)}

To investigate the expectations of SE researchers, we targeted members of the program committees of SE conferences. We designed and conducted an online survey to elicit their views on what aspects of PE they consider essential to report.
We selected a questionnaire-based approach using an online survey tool to obtain a broad and diverse set of responses, aiming for greater representativeness. Surveys are well established as effective means of gathering descriptive and retrospective insights, providing a valuable ``state-of-the-art overview on a particular method, tool, or technique''~\cite{punter2003conducting}.


The survey was carefully designed to enhance usability and participant engagement, as detailed below. Participants were guided through sequential pages, ensuring a clear, well-structured, and easy-to-navigate format. While online surveys do require a certain level of technological proficiency, we anticipated that our target demographic, i.e., SE researchers, would possess the necessary skills.

\subsection{Study Design}
\sectopic{Sampling of participants:} We targeted PC members from ICSE, ASE, and FSE for the years 2022--2024, which results in a total population of 612 potential participants. Given the international scope and nature of these conferences, the sample included researchers from diverse nationalities, backgrounds, and areas of expertise in SE research. Participants were contacted via email to request their participation in the study.

\sectopic{Survey design:} The survey was designed following the principles described by Kitchenham et al.~\cite{kitchenham2002principles}, and Punter et al.~\cite{punter2003conducting}. The full survey, including all questions, is included in our replication package. In the following, we describe the key aspects of the survey in more detail. While taking care to avoid overcrowding the screen with too many questions, we kept the number of pages manageable to reduce the risk of participants losing focus over time. The survey consisted of five pages with 2--4 questions each, along with an additional page containing a feedback text field. It was designed to take approximately 10 minutes to complete.

We included both \textit{closed-ended} and \textit{open-ended} questions in the survey. Closed-ended questions were used to collect clear and quantifiable data~\cite{punter2003conducting}. They provide straightforward answers (e.g., \textit{yes}, or \textit{no}), which simplifies both responding and subsequent analysis. In contrast, open-ended questions were included at the end of the survey to gain insight into additional contextual factors underlying participants' responses. These questions are essential for validating the data obtained from closed-ended questions and for gathering information that could not be captured otherwise. Efforts were made to minimize biases such as order effects, where responses to one question could influence answers to subsequent questions. The order of questions was carefully designed to mitigate this issue.

To encourage honest responses, anonymity was guaranteed to all participants. No personally identifiable information was collected. The survey remained open for a period of 30 days. To maximize the response rate, a reminder email was sent after three weeks to ask for participation from those who had not yet completed the survey.

\sectopic{Questionnaire content:} The final questionnaire comprised five content sections and one feedback section. The sections contained a total of 17 questions, including four open-ended questions (including one final feedback question) and 13 closed-ended questions. 
All sections, including the corresponding questions, are presented in \autoref{tab:survey-questions}. 
For the closed-ended questions, we employed the following scale, which resembles the one used in the ACM SIGSOFT Empirical Standards for Software Engineering~\cite{ralph2021empirical}:
\begin{compactitem}
    \item \textit{Essential:} A required element that \textit{must} be included in a paper to satisfy expectations for clarity, reproducibility, or rigor.
    \item \textit{Desirable:} A recommended element that \textit{should} be included to enhance quality, although it is not strictly necessary.
    \item \textit{Exceptional:} An advanced element that \textit{could} be included to go beyond typical expectations and significantly elevate the paper's overall quality.
    \item \textit{Not recommended:} An element that \textit{should not} be included in the paper, as it does not improve quality and may exceed the intended scope, introduce ambiguities, or cause other undesirable effects.
\end{compactitem}
We included definitions for all response options, both at the start of the questionnaire and at the top of each section, to enable easy reference throughout.

\begin{table}
    \centering
    \caption{Survey Questionnaire}
    \begin{tabularx}{\linewidth}{@{}l X@{}} \toprule
        \textbf{ID} & \textbf{Question} \\ \midrule
        \multicolumn{2}{@{}l@{}}{\textit{1. General Information}} \\ \midrule
        S1 & What is your primary area of expertise in SE? \\
        S2 & How familiar are you with the use of LLMs in SE research? \\
        S3 & How frequently do you review research papers involving LLMs? \\ \midrule
        \multicolumn{2}{@{}l@{}}{\textit{2. LLM Usage}} \\ \midrule
        S4 & Authors name the used LLMs. \\
        S5 & Authors precisely name the used LLM versions. \\
        S6 & Authors use different LLMs and compare the results. \\ \midrule
        \multicolumn{2}{@{}l@{}}{\textit{3. Prompt Usage}} \\ \midrule
        S7 & Authors describe the prompts used to solve a task. \\
        S8 & Authors provide the exact prompts used. \\
        S9 & Authors justify why a specific prompt structure or phrasing was chosen. \\
        S10 & Authors use and mention prompt engineering techniques to create prompts. \\ \midrule
        \multicolumn{2}{@{}l@{}}{\textit{4. Prompt Testing and Iterations}} \\ \midrule
        S11 & Authors report how they refined/iterated the prompts to improve performance. \\
        S12 & Authors apply automated prompt tuning techniques to optimize their prompts. \\
        S13 & Authors test multiple prompt variations and report the results. \\
        S14 & Authors discuss their use of prompts as part of threats to validity or potential limitations. \\ \midrule
        \multicolumn{2}{@{}l@{}}{\textit{5. Overlooked Aspects of Prompting}} \\ \midrule
        S15 & Are there any aspects of prompt usage or documentation that you feel are often overlooked in research papers? \\
        S16 & Is there another aspect or comment you would like to add regarding prompt usage and documentation? \\ \midrule
        \multicolumn{2}{@{}l@{}}{\textit{6. Feedback}} \\ \midrule
        S17 & Do you have any comments, suggestions, or feedback about this survey? \\
        \bottomrule
    \end{tabularx}
    \label{tab:survey-questions}
\end{table}

\subsection{Study Results}

We contacted 612 former PC members, of whom 105 (17.16~\%) responded. A total of 92 responses were complete, with all closed-ended questions answered. This corresponds to a response rate of 15.03~\%, which is considerably higher than the anticipated rate of 5~\% typically achieved in questionnaire-based SE surveys~\cite{singer2008software}. The complete dataset is available in our replication package.

Among the 92 complete responses, 59 participants (64.13~\%) reported being familiar with LLMs, while 31 (33.7~\%) were somewhat familiar. Only 2 participants (2.17~\%) indicated having no familiarity with LLMs and were therefore excluded from the results.

\begin{figure}
    \centering
    \includegraphics[width=\linewidth]{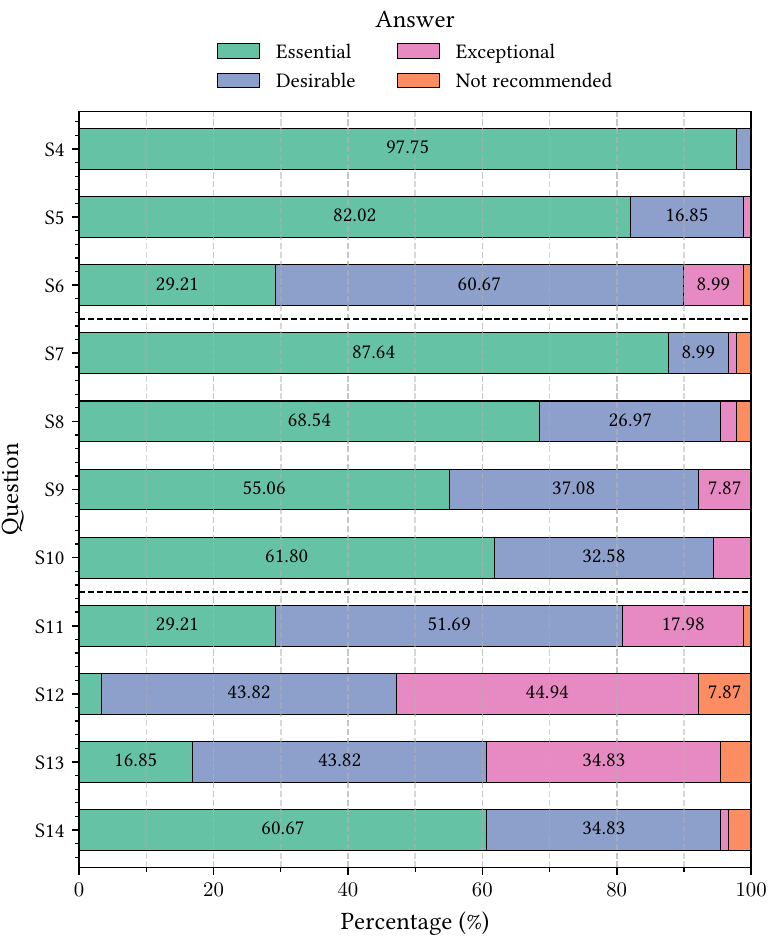}
    \caption{A bar chart showing the percentages of closed-ended survey questions (cf. \autoref{tab:survey-questions}) answered with either \textit{Essential}, \textit{Desirable}, \textit{Exceptional}, or \textit{Not recommended}.}
    \label{fig:survey-closed-questions}
\end{figure}

When asked about their experience reviewing research papers involving LLMs, a majority of participants (70; 76.09~\%) reported reviewing such papers frequently (more than 5 papers per year), while 19 (20.65~\%) indicated doing so occasionally (1--4 papers per year). Only 3 participants (3.26~\%) stated that they have never reviewed such papers so far. Their responses were excluded from the analysis to ensure the dataset reflected participants with relevant experience. Of these participants, 2 were the same individuals who reported having no familiarity with LLMs, resulting in 89 responses for analysis. The results of all closed-questions can be seen in \autoref{fig:survey-closed-questions}.

\sectopic{LLM usage:} Most participants (87 out of 89; 97.75~\%) agreed that naming the LLMs used is essential, with a majority (82.02~\%) emphasizing the importance of specifying the exact version. Some respondents stressed this point also in the free-text responses, especially concerning the fast-developing landscape of LLMs (e.g., ``\textit{Prompts’ effectiveness may depend on the parameters of the LLM, which evolve fast.}'', ``\textit{LLMs are evolving so prompt engineering techniques are also in flux.}''). 
Furthermore, 60.67~\% of the participants considered comparing results across different LLMs to be desirable, while 29.21~\% regarded it as essential.

\sectopic{Prompt usage:} The majority of participants (87.65~\%) supported describing prompts, and most (68.54~\%) considered including the full prompt to be essential. Providing exact prompts was the most frequently mentioned concern, appearing in multiple answers across the free-text answer fields (e.g., ``\textit{I rarely see exact prompts used, which I guess is understandable given page limits, but still disappointing.}''). More than half of the participants (55.06~\%) indicated that providing sufficient justification for how prompts are constructed is essential. Additionally, 61.8~\% of participants regarded utilizing and reporting on PE techniques as essential, while 32.58~\% considered it desirable.

\sectopic{Prompt testing and iterations:} The aspects covered in this section were generally perceived as elements that enhance research quality rather than being strictly necessary. Approximately half of the participants (51.69~\%) considered refining or iterating on prompts to be desirable, while 29.21~\% deemed it essential. However, in the free-text fields, several respondents felt that authors often do not explain how prompts were developed, tuned, or selected (e.g., ``\textit{Yes, the details of concrete prompts and prompt tuning strategies are often lacking.}'', ``\textit{One needs a sort of pre-experiment to find the right prompt.}'').

Prompt tuning was rated as desirable by 43.82~\% and as exceptional by 44.94~\%. This question also received the highest number of \textit{not recommended} responses (7; 7.87~\%). In the comments, participants raised concerns about the risk of overfitting prompts to specific LLMs when applying prompt tuning (e.g., ``\textit{Overfitting of prompts to specific LLMs is starting to be a big issue.}''). The use of multiple prompt variations was largely seen as beneficial but not mandatory, with 43.82~\% regarding it as essential and 34.83~\% as exceptional.

Furthermore, the majority of participants (60.67~\%) believed that acknowledging prompt-related threats to validity is essential, while 34.83~\% considered it desirable.

\sectopic{Overlooked aspects of prompting:} In this free-text response field, participants highlighted the need for clearer documentation of LLM settings and configurations. Five participants emphasized that LLM parameters (e.g., temperature and top-p) should be reported alongside prompts and datasets, as all these factors influence model behavior and are important for reproducibility. Six participants noted the absence of a rationale for model selection. Additionally, four participants pointed out the lack of information on computational costs, including resource consumption and financial expenses associated with running LLMs on commercial APIs.

Generally, the importance of reproducibility was emphasized. One participant suggested that commercial models should not be solely relied upon due to concerns about their long-term availability. Another participant argued that prompt creation should be viewed as a form of program development, suggesting that the entire lifecycle, from design to development and testing, should be systematically documented.

\subsection{Threats to Validity}

\sectopic{Construct validity:}
A key threat to construct validity is the potential misinterpretation of survey items by participants. To reduce ambiguity, we used standard SE terminology and clearly defined the rating scale (i.e., \textit{Essential}, \textit{Desirable}, \textit{Exceptional}, \textit{Not Recommended}). However, differences in individual interpretation of these categories may still affect consistency across responses.  
To mitigate wording bias, we phrased items as neutral statements (e.g., \textit{``Authors name the used LLMs''}) rather than value-laden questions (e.g., \textit{``Is it essential to name the LLM?''}). Despite these efforts, subtle bias in how items were framed may still have influenced responses.  
The granularity of our four-point scale also limits expressiveness: some participants may have found it difficult to fully express nuanced opinions within these fixed categories.  
Finally, the selection of reporting items itself may introduce bias, as the list was derived from our literature review (Section~\ref{sec:rq1}) and may not include all dimensions that participants consider relevant.

\sectopic{Internal validity:} 
Surveys inherently lack interactivity, which limits opportunities for clarification or follow-up. Unlike interviews, we could not probe deeper into ambiguous or contradictory answers. This constraint was accepted as a trade-off to support quantitative analysis and ensure consistency with the literature review.  
Additionally, self-reporting bias may be present, as participants might have responded in ways they perceived as socially or academically acceptable, rather than fully reflecting their typical reviewing practices.

\sectopic{External validity:}
Our survey targeted researchers who have served on program committees for top SE conferences. While this group was appropriate for assessing reviewer expectations, their views may not fully align with those of developers, practitioners, or researchers in other domains actively working with LLMs.  
This academic focus may bias the results toward expectations grounded in scientific transparency rather than industrial pragmatism. Furthermore, non-response bias is a concern: participants with a strong interest in LLMs or prompting may have been more likely to respond. This could overemphasize the importance of prompt reporting. However, based on open-ended responses and critique diversity, we observed participation from both proponents and skeptics of LLM-based research, suggesting a range of perspectives was represented.

\section{Alignment of Current State with Expectations (RQ3)}
For RQ3, we compare the extracted practices with the expectations assessed in our survey.  
For this purpose, we first derive a guideline from the survey responses and then compare it with the reporting practices identified in our review of the literature.

\subsection{Guideline Derivation}
We derive a guideline by analyzing the perceived importance of different reporting elements collected through our survey. 
We used statistical methods to analyze the differences in the response patterns between the survey items. We first used the Friedman test (a non-parametric omnibus test for repeated measures) to test whether there are statistically significant differences between the response patterns to survey items. The Friedman test yielded a highly significant result ($p < 0.000001$ with $\alpha=0.05$), indicating strong evidence that there are differences in responses in the criteria. Hence, we reject the null hypothesis that all criteria share the same response distribution.
To identify these differences, we conducted pairwise Wilcoxon signed-rank tests as post-hoc tests for all pairs of criteria. The unadjusted p-values were corrected with the Bonferroni procedure.

\begin{figure}
    \centering
    \includegraphics[width=1\columnwidth]{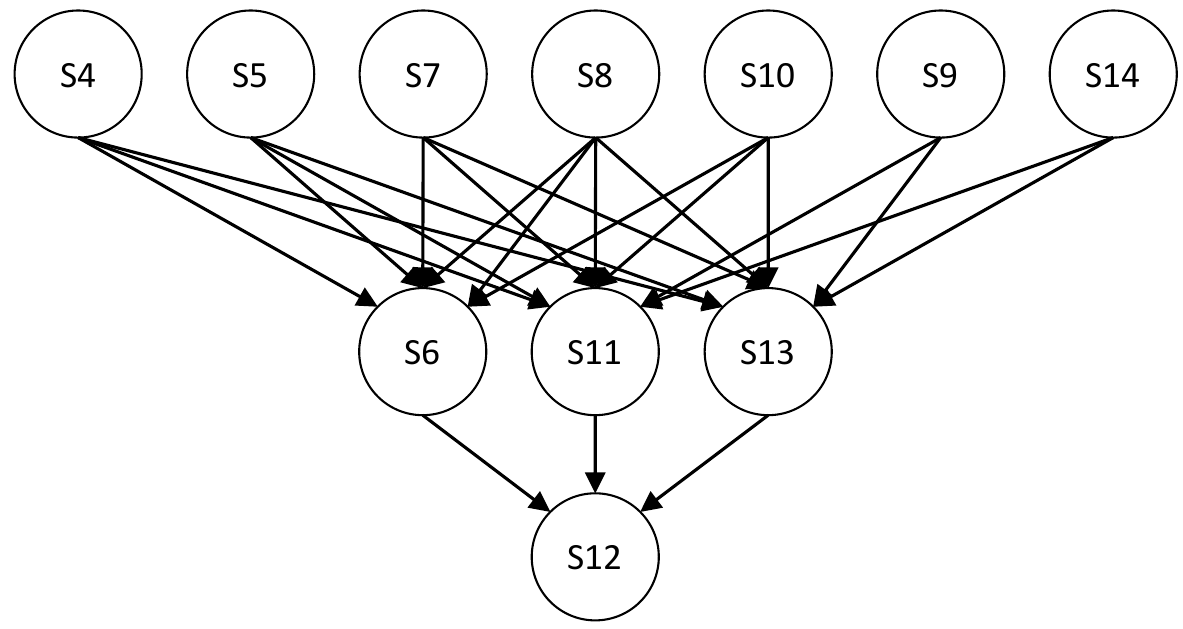}
    \caption{Survey items (S4--S14) and their response patterns. An arrow pointing from node $A$ to $B$ indicates that $A$ had a significantly higher median response than $B$.}
    \label{fig:guideline-graph}
\end{figure}

Figure~\ref{fig:guideline-graph} shows a graph-based representation of the significant differences between the response patterns of the survey items (S4--S14). Arrows indicate statistically significant pairwise differences between response items (Wilcoxon signed-rank test, Bonferroni-corrected, $\alpha=0.05$).

Based on this analysis, three groups of items emerged, which we characterized as \textit{essential}, \textit{desirable}, and \textit{exceptional} elements. This classification aligns with the ACM SIGSOFT Empirical Standards~\cite{ralph2021empirical}. To formulate the guideline, we followed the idea of Baltes~et~al.~\cite{baltes2025guidelines} and used \textit{must}, \textit{should}, and \textit{may} as suggested in RFC~2119\footnote{\url{https://www.rfc-editor.org/rfc/rfc2119}}.
We assigned items to essential, desirable, and exceptional groups based on their median response ranks and statistically significant differences identified through post-hoc comparisons.

Our final guidelines are shown in~\ref{tbl:guideline}, categorized into three groups depending on their importance. The table outlines the key reporting elements along with their classification and reported literature frequencies, grounding our recommendations on expert judgment and comparing them to empirical evidence. The values in column ``PC member agreement'' correspond to the ratio of respondents who categorized the item in the respective group (e.g., 44.94~\% of respondents categorized the use of automated prompt tuning techniques as exceptional).

\begin{figure}
    \centering
    \includegraphics[width=.8\linewidth]{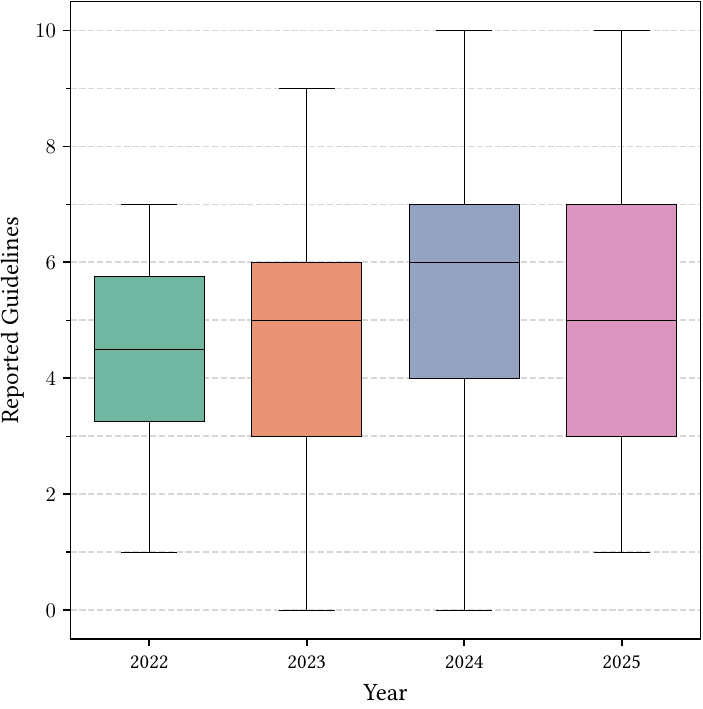}
    \caption{A box plot showing, for each year of extracted papers, the number of reported guidelines per paper.}
    \label{fig:reported-guidelines-boxplot}
\end{figure}

To assess whether reporting practices have improved over time, we analyzed trends in guideline adherence across publication years. \autoref{fig:reported-guidelines-boxplot} shows the distribution of guideline items followed per paper by year. The data suggests a slight upward trend in adherence, though the median remains modest, ranging from 4.5 to 6 out of 11 possible items. The variance is also considerable: while some papers follow nearly all recommended practices, a significant number report only three or fewer, highlighting continued inconsistency in reporting standards.

\begin{table*}
    \centering
    \caption{Guidelines on Reporting LLM Prompting}
    \begin{tabularx}{\textwidth}{@{}X r r@{}}
        \toprule
        \multirow{2}{*}{\textbf{Guideline}} & \textbf{Followed} & \textbf{PC Member} \\
        & \textbf{in Papers} & \textbf{Agreement} \\ \midrule
        \multicolumn{2}{@{}l@{}}{\textit{Essential}} \\ \midrule
        Authors \textit{must} name the used LLMs (e.g., \textit{GPT-4}, \textit{Llama 3}, \textit{Claude Opus 4}). & 92.31~\% & 97.75~\% \\
        Authors \textit{must} precisely name the used LLM versions (e.g., \textit{GPT-4 2024-08-06}). & 16.43~\% & 82.02~\% \\
        Authors \textit{must} provide the exact prompts used word-by-word. They may shorten the prompt using templates. & 57.69~\% & 68.54~\% \\
        Authors \textit{must} describe the prompts used and how they are structured. & 66.08~\% & 87.64~\% \\
        Authors \textit{must} justify why a specific prompt structure or phrasing was chosen. & 51.40~\% & 55.06~\% \\
        Authors \textit{must} mention all prompt engineering techniques used (e.g., \textit{few-shot}, \textit{chain-of-thought}). & 62.24~\% & 61.80~\% \\
        Authors \textit{must} discuss their use of prompts as part of threats to validity. & 21.68~\% & 60.67~\% \\ \midrule
        \multicolumn{2}{@{}l@{}}{\textit{Desirable}} \\ \midrule
        Authors \textit{should} use different LLMs and compare the results. & 56.99~\% & 60.67~\% \\
        Authors \textit{should} report how they refined/iterated the prompts to improve performance. & 45.45~\% & 51.69~\% \\
        Authors \textit{should} test multiple prompt variations and report the results. & 43.36~\% & 43.82~\% \\ \midrule
        \multicolumn{2}{@{}l@{}}{\textit{Exceptional}} \\ \midrule
        Authors \textit{may} apply automated prompt tuning techniques. & 4.90~\% & 44.94~\% \\
        \bottomrule
    \end{tabularx}
    \label{tbl:guideline}
\end{table*}

\subsection{Discussion of Practices}

\sectopic{Quantitative comparison of expectations and practices:}
The results reveal notable gaps between community expectations and current reporting practices. In particular, two guideline items exhibit especially large discrepancies. First, only 16.4~\% of the analyzed papers reported the exact version of the LLM used, despite more than 80~\% of survey respondents identifying this as an essential practice. This gap may be due to limited awareness among authors, e.g., some may not realize that models like GPT-4 are regularly updated even under the same version name, or it may reflect a belief that such version differences are negligible.

Second, only around 20~\% of papers addressed prompting as a potential threat to validity, whereas over 60~\% of respondents considered this discussion essential. This mismatch may be attributed to the relative novelty of prompting in SE research, where community norms around validity threats and mitigation strategies have yet to be established.

While the proportion of papers following the desirable and exceptional items generally aligns with reviewer expectations, adherence to essential items remains inconsistent. Most essential items were reported in only 51–66~\% of papers, indicating substantial room for improvement.

\sectopic{Interpretation and implications:}
There are several possible reasons for the mismatch between how often papers follow the proposed guideline items and how strongly reviewers expect them to be reported. A straightforward explanation is the strict page limits imposed by major SE conferences, which often force authors to omit methodological details perceived as less important. This effect may be amplified by the lack of community consensus on whether prompts constitute part of the method or merely the experimental setup. Many researchers still treat prompt wording as an implementation detail rather than as a methodological decision that impacts the results and reproducibility.

Reviewer bias may further reinforce this cycle. For example, if only about two-thirds of reviewers consider including the exact prompt as essential, the remaining third may overlook its omission during review. As a result, authors receive inconsistent feedback and may lower the priority of prompt reporting in subsequent submissions. Over time, such variability in reviewing standards can lead to more brief reporting practices, even when most reviewers conceptually value transparency.

Equally important as understanding why these mismatches occur is recognizing how they affect the replicability and methodological soundness of published work. With the rapid evolution of current LLMs, omitting the exact version used in experiments already severely limits replicability as different updates of the same model may produce noticeably different outputs. Likewise, not reporting the exact prompt can hinder reproducibility entirely, as subsequent researchers cannot reconstruct the original interaction that generated the results.

Methodological soundness further depends on transparent reasoning behind experimental decisions. When authors neither justify why specific prompts were chosen nor reviewers consistently request such explanations, the conceptual foundation of a study becomes weaker. The fact that most papers omit prompting as a potential threat to validity illustrates that prompting is still not widely regarded as a methodological factor that can bias outcomes if not applied carefully. Addressing these reporting omissions is therefore essential to ensure credible, reproducible, and theoretically grounded LLM-based SE research.

To help close these gaps, we propose that authors adopt prompt-reporting templates that facilitate the inclusion of longer prompts and detailed descriptions without exceeding page limits. Authors should also critically evaluate which LLM-specific details, such as model version, system settings, or fine-tuning parameters, are necessary to maximize reproducibility and methodological soundness. Reviewers, in turn, could consider checklists to ensure completeness of such information during evaluation. Integrating these empirically grounded standards directly into review forms would simplify the review process while reinforcing consistent expectations.

\section{Conclusions and Future Directions}

\sectopic{General limitations of our approach:}
While our guideline is grounded in a systematic literature review and a survey of experienced SE researchers, the methodology carries several general limitations that should be acknowledged.

First, expectations and practices are evolving rapidly in the domain of LLM-based software engineering. As prompting techniques, LLM capabilities, and community norms continue to develop, parts of the proposed guideline may become outdated or require revision.

Second, our approach may introduce a descriptive versus prescriptive bias. The literature review reflects what authors chose to report, which may not always correspond to best practices. Similarly, survey responses indicate what participants believe should be reported, which may be influenced by individual experience, norms, or exposure rather than empirical validation of reporting effectiveness.

Third, the guideline may overgeneralize across diverse use cases. Prompting strategies and documentation needs vary across SE tasks (e.g., code generation vs. test synthesis), LLM configurations (e.g., fine-tuned models vs. zero-shot APIs), and study types~\cite{baltes2025guidelines}. A single unified set of recommendations may not fully account for these task- and context-specific differences. This was also highlighted by some of our respondents (e.g., ``\textit{For me, a lot of this depends on the RQs. If the prompt is core to the experiment, it must be disclosed.}'')

Finally, the proposed guideline has not yet been empirically validated in terms of its practical impact. While we believe it can improve reproducibility and transparency, future work is needed to assess whether guideline adoption leads to measurable improvements in review quality, replicability, or research clarity.

These limitations underscore the need to treat our guideline as an empirically grounded starting point, rather than a fixed or universal standard, and to revisit and refine it as the field matures.

\sectopic{Relation to existing evidence:}
Our guideline overlaps with Baltes et al.'s~\cite{baltes2025guidelines} in key areas, such as the importance of reporting exact prompts, describing their structure, and documenting prompt engineering techniques. We extend their work by providing quantitative data on how often practices are followed and to what extent reviewers expect them, thus offering an evidence-based prioritization.
Importantly, we do not claim to replace or supersede the broader LLM guidelines proposed by Baltes et al. Rather, we view our work as a complementary effort, focused specifically on prompt reporting within SE research, and as an empirical validation of key prompt-related aspects from their broader proposal. Where Baltes et al. provide an expert-driven vision, our work aims to anchor that vision in current practice and reviewer expectations.

\sectopic{Future work:}
Our study opens up several directions for future research and community engagement. 
One promising avenue is to expand the current guidelines toward the emerging paradigms of ``promptware''~\cite{ChenEtAl2025} and ``AI-ware''\footnote{\url{https://conf.researchr.org/home/aiware-2025}}, where prompts become persistent artifacts embedded in software systems. In these contexts, documenting prompts is critical not just for research reproducibility but also for long-term maintainability, debugging, and managing technical debt. Alongside prompt engineering, \textit{context engineering} is gaining importance as a technique for managing information within the limited context window of LLMs. Future guidelines could incorporate documentation practices for context segmentation, token compression, and retrieval-augmented generation, which are increasingly relevant in complex LLM-powered systems.

Another important direction is the empirical validation of our reporting guidelines. While our survey and literature analysis support their relevance, future work could assess their impact on actual research quality (e.g., by measuring improvements in reproducibility, peer review scores, or clarity of experimental design). Moreover, while our study focused on the software engineering domain, similar prompting practices are used in other fields such as HCI, data science, and NLP. Investigating how these guidelines transfer to or need adaptation in other research communities would help ensure their broader applicability.

To promote practical adoption, tooling and author/reviewer support could be developed. This may include prompt reporting templates, automated checklists, or integration with artifact evaluation processes. Given the pace of innovation in LLM research, we also envision maintaining the guidelines as a living resource, allowing them to evolve in response to emerging tools, prompting strategies, and community norms.

Finally, we aim to contribute our findings to broader community standardization efforts. In particular, we see opportunities to collaborate with the complementary work by Baltes et al.~\cite{baltes2025guidelines}, whose top-down guidelines address a broader range of LLM-related research practices. Our empirically grounded, bottom-up results provide a valuable counterpoint and validation. We also plan to offer our findings as input to the ACM SIGSOFT Empirical Standards\footnote{\url{https://www2.sigsoft.org/EmpiricalStandards/}}, which currently lack guidance on LLM-driven research. By contributing to these community-driven initiatives, we hope to support the development of consistent, high-quality practices for documenting and evaluating LLM usage in software engineering and beyond.

\section*{Acknowledgments}
We used Generative AI (e.g., Gemini-2.5 and ChatGPT) to support and validate data extraction and filtering, and to improve the text.


\bibliographystyle{ACM-Reference-Format}
\bibliography{references}

@incollection{singer2008software,
  title={Software engineering data collection for field studies},
  author={Singer, Janice and Sim, Susan E and Lethbridge, Timothy C},
  booktitle={Guide to advanced empirical software engineering},
  pages={9--34},
  year={2008},
  publisher={Springer}
}

@misc{nguyen2023generative,
  doi = {10.48550/ARXIV.2310.18648},
  url = {https://arxiv.org/abs/2310.18648},
  author = {Nguyen-Duc,  Anh and Cabrero-Daniel,  Beatriz and Przybylek,  Adam and Arora,  Chetan and Khanna,  Dron and Herda,  Tomas and Rafiq,  Usman and Melegati,  Jorge and Guerra,  Eduardo and Kemell,  Kai-Kristian and Saari,  Mika and Zhang,  Zheying and Le,  Huy and Quan,  Tho and Abrahamsson,  Pekka},
  title = {Generative Artificial Intelligence for Software Engineering -- A Research Agenda},
  publisher = {arXiv},
  year = {2023},
  copyright = {Creative Commons Attribution 4.0 International}
}

@inproceedings{Shin2025,
  title = {Prompt Engineering or Fine-Tuning: An Empirical Assessment of {LLMs} for Code},
  DOI = {10.1109/msr66628.2025.00082},
  booktitle = {IEEE/ACM 22nd International Conference on Mining Software Repositories (MSR)},
  publisher = {IEEE},
  author = {Shin,  Jiho and Tang,  Clark and Mohati,  Tahmineh and Nayebi,  Maleknaz and Wang,  Song and Hemmati,  Hadi},
  year = {2025},
  pages = {490--502}
}

@misc{Sahoo24,
  doi = {10.48550/ARXIV.2402.07927},
  author = {Sahoo,  Pranab and Singh,  Ayush Kumar and Saha,  Sriparna and Jain,  Vinija and Mondal,  Samrat and Chadha,  Aman},
  title = {A Systematic Survey of Prompt Engineering in Large Language Models: Techniques and Applications},
  publisher = {arXiv},
  year = {2024},
  copyright = {Creative Commons Attribution 4.0 International}
}

@inproceedings{brown2020language,
	title        = {Language {{Models}} Are {{Few-Shot Learners}}},
	author       = {Brown, Tom and Mann, Benjamin and Ryder, Nick and Subbiah, Melanie and Kaplan, Jared D and Dhariwal, Prafulla and Neelakantan, Arvind and Shyam, Pranav and Sastry, Girish and Askell, Amanda and Agarwal, Sandhini and {Herbert-Voss}, Ariel and Krueger, Gretchen and Henighan, Tom and Child, Rewon and Ramesh, Aditya and Ziegler, Daniel and Wu, Jeffrey and Winter, Clemens and Hesse, Chris and Chen, Mark and Sigler, Eric and Litwin, Mateusz and Gray, Scott and Chess, Benjamin and Clark, Jack and Berner, Christopher and McCandlish, Sam and Radford, Alec and Sutskever, Ilya and Amodei, Dario},
	year         = {2020},
	booktitle    = {Advances in {{Neural Information Processing Systems}}},
	publisher    = {Curran Associates, Inc.},
	volume       = {33},
	pages        = {1877--1901},
	urldate      = {2025-06-17}
}

@article{hou2024large,
	title        = {Large {{Language Models}} for {{Software Engineering}}: {{A Systematic Literature Review}}},
	shorttitle   = {Large {{Language Models}} for {{Software Engineering}}},
	author       = {Hou, Xinyi and Zhao, Yanjie and Liu, Yue and Yang, Zhou and Wang, Kailong and Li, Li and Luo, Xiapu and Lo, David and Grundy, John and Wang, Haoyu},
	year         = {2024},
	month        = dec,
	journal      = {ACM Trans. Softw. Eng. Methodol.},
	volume       = {33},
	number       = {8},
	pages        = {220:1--220:79},
	doi          = {10.1145/3695988},
	issn         = {1049-331X},
	urldate      = {2025-06-17}
}

@article{liu2023pre-train,
	title        = {Pre-Train, {{Prompt}}, and {{Predict}}: {{A Systematic Survey}} of {{Prompting Methods}} in {{Natural Language Processing}}},
	shorttitle   = {Pre-Train, {{Prompt}}, and {{Predict}}},
	author       = {Liu, Pengfei and Yuan, Weizhe and Fu, Jinlan and Jiang, Zhengbao and Hayashi, Hiroaki and Neubig, Graham},
	year         = {2023},
	month        = jan,
	journal      = {ACM Comput. Surv.},
	volume       = {55},
	number       = {9},
	pages        = {195:1--195:35},
	doi          = {10.1145/3560815},
	issn         = {0360-0300},
	urldate      = {2025-06-17}
}

@inproceedings{punter2003conducting,
  title = {Conducting on-line surveys in software engineering},
  DOI = {10.1109/isese.2003.1237967},
  booktitle = {International Symposium on Empirical Software Engineering (ISESE)},
  publisher = {IEEE},
  author = {Punter,  T. and Ciolkowski,  M. and Freimut,  B. and John,  I.},
  year = {2003},
  pages = {80--88}
}

@misc{ralph2021empirical,
  title = {Empirical {{Standards}} for {{Software Engineering Research}}},
  author = {Ralph, Paul and bin Ali, Nauman and Baltes, Sebastian and Bianculli, Domenico and Diaz, Jessica and Dittrich, Yvonne and Ernst, Neil and Felderer, Michael and Feldt, Robert and Filieri, Antonio and de Fran{\c c}a, Breno Bernard Nicolau and Furia, Carlo Alberto and Gay, Greg and Gold, Nicolas and Graziotin, Daniel and He, Pinjia and Hoda, Rashina and Juristo, Natalia and Kitchenham, Barbara and Lenarduzzi, Valentina and Mart{\'i}nez, Jorge and Melegati, Jorge and Mendez, Daniel and Menzies, Tim and Molleri, Jefferson and Pfahl, Dietmar and Robbes, Romain and Russo, Daniel and Saarim{\"a}ki, Nyyti and Sarro, Federica and Taibi, Davide and Siegmund, Janet and Spinellis, Diomidis and Staron, Miroslaw and Stol, Klaas and Storey, Margaret-Anne and Taibi, Davide and Tamburri, Damian and Torchiano, Marco and Treude, Christoph and Turhan, Burak and Wang, Xiaofeng and Vegas, Sira},
  year = {2021},
  month = mar,
  number = {arXiv:2010.03525},
  eprint = {2010.03525},
  primaryclass = {cs},
  publisher = {arXiv},
  doi = {10.48550/arXiv.2010.03525},
  urldate = {2025-07-04},
  archiveprefix = {arXiv}
}

@article{kitchenham2002principles,
  title = {Principles of Survey Research Part 2: Designing a Survey},
  shorttitle = {Principles of Survey Research Part 2},
  author = {Kitchenham, Barbara A. and Pfleeger, Shari Lawrence},
  year = {2002},
  month = jan,
  journal = {SIGSOFT Softw. Eng. Notes},
  volume = {27},
  number = {1},
  pages = {18--20},
  issn = {0163-5948},
  doi = {10.1145/566493.566495},
  urldate = {2025-07-04}
}

@article{kitchenham2007guidelines,
  title={Guidelines for performing systematic literature reviews in software engineering},
  author={Kitchenham, Barbara and Charters, Stuart and others},
  year={2007},
  publisher={Keele, UK}
}

@misc{baltes2025guidelines,
      title={Guidelines for Empirical Studies in Software Engineering involving Large Language Models}, 
      author={Sebastian Baltes and Florian Angermeir and Chetan Arora and Marvin Muñoz Barón and Chunyang Chen and Lukas Böhme and Fabio Calefato and Neil Ernst and Davide Falessi and Brian Fitzgerald and Davide Fucci and Marcos Kalinowski and Stefano Lambiase and Daniel Russo and Mircea Lungu and Lutz Prechelt and Paul Ralph and Rijnard van Tonder and Christoph Treude and Stefan Wagner},
      year={2025},
      eprint={2508.15503},
      archivePrefix={arXiv},
      primaryClass={cs.SE},
      url={https://arxiv.org/abs/2508.15503}, 
}

@misc{ChenEtAl2025,
  doi = {10.48550/ARXIV.2503.02400},
  url = {https://arxiv.org/abs/2503.02400},
  author = {Chen,  Zhenpeng and Wang,  Chong and Sun,  Weisong and Yang,  Guang and Liu,  Xuanzhe and Zhang,  Jie M. and Liu,  Yang},
  title = {Promptware Engineering: Software Engineering for {LLM} Prompt Development},
  publisher = {arXiv},
  year = {2025},
  copyright = {Creative Commons Attribution 4.0 International}
}

@misc{TrinkenreichEtAl2025,
  doi = {10.48550/ARXIV.2506.12691},
  url = {https://arxiv.org/abs/2506.12691},
  author = {Trinkenreich,  Bianca and Calefato,  Fabio and Hanssen,  Geir and Blincoe,  Kelly and Kalinowski,  Marcos and Pezzè,  Mauro and Tell,  Paolo and Storey,  Margaret-Anne},
  title = {Get on the Train or be Left on the Station: Using {LLMs} for Software Engineering Research},
  publisher = {arXiv},
  year = {2025},
  copyright = {Creative Commons Attribution 4.0 International}
}

@misc{huang2025,
      title={Prompt Engineering for Requirements Engineering: A Literature Review and Roadmap}, 
      author={Kaicheng Huang and Fanyu Wang and Yutan Huang and Chetan Arora},
      year={2025},
      eprint={2507.07682},
      archivePrefix={arXiv},
      primaryClass={cs.SE},
      url={https://arxiv.org/abs/2507.07682}, 
}

@inproceedings{ZanCZLWGWL23,
  title = {Large Language Models Meet {NL2Code}: A Survey},
  DOI = {10.18653/v1/2023.acl-long.411},
  booktitle = {61st Annual Meeting of the Association for Computational Linguistics (Volume 1: Long Papers)},
  publisher = {Association for Computational Linguistics},
  author = {Zan,  Daoguang and Chen,  Bei and Zhang,  Fengji and Lu,  Dianjie and Wu,  Bingchao and Guan,  Bei and Yongji,  Wang and Lou,  Jian-Guang},
  year = {2023}
}

@article{ZhangFMSC24,
  author       = {Quanjun Zhang and
                  Chunrong Fang and
                  Yuxiang Ma and
                  Weisong Sun and
                  Zhenyu Chen},
  title        = {A Survey of Learning-based Automated Program Repair},
  journal      = {{ACM} Trans. Softw. Eng. Methodol.},
  volume       = {33},
  number       = {2},
  pages        = {55:1--55:69},
  year         = {2024},
  url          = {https://doi.org/10.1145/3631974},
  doi          = {10.1145/3631974},
  bibsource    = {dblp computer science bibliography, https://dblp.org}
}

@article{WangHCLWW24,
  author       = {Junjie Wang and
                  Yuchao Huang and
                  Chunyang Chen and
                  Zhe Liu and
                  Song Wang and
                  Qing Wang},
  title        = {Software Testing With Large Language Models: Survey, Landscape, and
                  Vision},
  journal      = {{IEEE} Trans. Software Eng.},
  volume       = {50},
  number       = {4},
  pages        = {911--936},
  year         = {2024},
  url          = {https://doi.org/10.1109/TSE.2024.3368208},
  doi          = {10.1109/TSE.2024.3368208},
  bibsource    = {dblp computer science bibliography, https://dblp.org}
}

\end{document}